\documentclass[prl,showpacs,twocolumn]{revtex4}  
\usepackage{graphics}

\begin{document}

\title{Achieving peak brightness in an atom laser}
\author{N. P. Robins}
\author{C. Figl}
\author{S.A. Haine}
\author{A. K. Morrison}
\author{M. Jeppesen}
\author{J. J. Hope}
\author{J. D. Close}

\affiliation{Australian Centre for Quantum Atom Optics, Physics Department, The Australian National University, Canberra, 0200, Australia.}
\email{nick.robins@anu.edu.au}
\homepage{http://www.acqao.org}

\begin{abstract}
In this paper we present experimental results and theory on the first continuous (long pulse) Raman atom laser. The brightness that can be achieved with this system is three orders of magnitude greater than has been previously demonstrated in any other continuously outcoupled atom laser. In addition, the energy linewidth of a continuous atom laser can be made arbitrarily narrow compared to the mean field energy of a trapped condensate.  We analyze the flux and brightness of the atom laser with an analytic model that shows excellent agreement with experiment with no adjustable parameters.  
\end{abstract}


\maketitle

The analogy between atom lasers and optical lasers is strong \cite{wise}. Both optical and atom lasers create a coherent output beam of bosons that are photons in the case of an optical laser and deBroglie matter waves in the case of an atom laser. The lasing mode in an optical laser is pumped through a non-thermal equilibrium process and is not the lowest energy mode of the cavity but, rather, is a highly excited mode many wavelengths long. For atoms, the lasing mode is populated through Bose-Einstein condensation and is the ground state of the trap.  In an optical laser, a beam is outcoupled using a grating or a patially reflecting mirror. Most atom laser beams studied to date have been outcoupled either continuously (long pulse) or pulsed using RF radiation to transfer atoms from a trapped magnetic sub-state to a state that does not interact with the trapping field \cite{mewes,bloch,aspect}. The atoms  fall away from the trap under gravity producing a coherent matter wave. 

\begin{figure}[b]
\centerline{\scalebox{.55}{\includegraphics{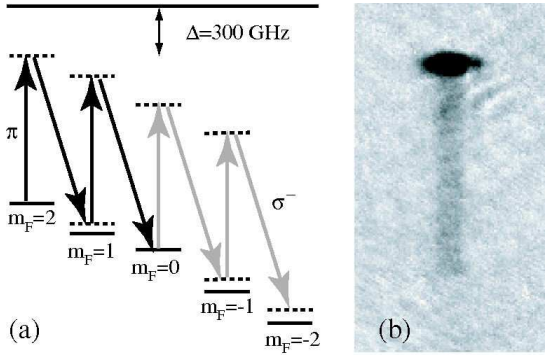}}}
\caption{A continuous Raman output-coupler. (a) Transitions from the  ${\rm 5^2 S_\frac{1}{2} F = 2, m_F=2}$ state to the ${\rm F = 2, m_F=0}$ state via the $\rm{5^2 P_\frac{3}{2}}$ transition of $\rm{^{87}Rb}$.  (b) Absorption image of an 8.5\,ms continuous Raman atom laser produced with 60\,$\mu$W per 1/,mm beam.  }
\end{figure}

Optical lasers have found broad application in precision measurements to address questions both fundamental and applied in nature.  In many cases, we expect to be able to perform such experiments more effectively and to higher precision with atom interferometry \cite{berman}.  Measurements made with the low density, highly coherent beam from an atom laser will not be limited in precision by the mean field interaction that plagues interferomteric measurements made with more dense atomic sources such as full Bose Einstein condensates. The development of the atom laser past the demonstration stage to a useful tool is an important goal. 

In principle, in a precision interferometric measurement made at the shot noise limit, all that is required of the wave source, whether it be a source of matter waves or light, is that it have high flux and that the normalised second order correlation function, ${\rm g_2(\tau) = 1}$ \cite{ottl}.  In principle, classical source fluctuations, both in frequency and phase, can be removed through good interferometer design, a fact well known to the gravity wave community \cite{david}.  Long coherence length, equivalent to a spectrally narrow source, is not required if the path length difference in the interferometer is less than the coherence length.  In principle, mode matching on the output beam splitter of an interferometer can be performed as well on a complicated spatial mode as a simple one and a highly divergent  beam can be collimated with lenses. In practice, if an interferometer is to operate at the shot noise limit, none of this is true. The shot noise limit is a difficult limit to achieve at least for high flux, and it is essential to have a spectrally narrow, classically quiet, low divergence beam with the minimum transverse structure in both phase and amplitude. For these practical reasons, classically quiet lasers operating on the TEM$_{00}$ mode are used in precision optical measurements in order that we can, in a real experiment, reach the quantum noise limit.  The same reasoning applies to atom lasers. 

In this paper, we present the first results on a continuously outcoupled Raman atom laser.  Raman out-coupling of an atom laser beam from a magnetically trapped condensate, achieved with two phase-locked optical lasers, flips atoms to a non-trapped state and provides a momentum kick equal to twice the momentum per photon of the optical outcoupling beams (see figure 1). Prior to the work presented here, there has been one demonstration of a Raman atom laser by Hagley {\em et al}. that was performed in pulsed mode producing a broad-linewidth (${\rm \Delta E\sim}$1\,kHz) multi-spin component beam \cite{hagley}.   A continuous output coupler based on the Bragg effect, a non state changing two photon transition, has recently been utilized to demonstrate the formation of the relative phase between BECs in two separate optical traps \cite{saba}.  Our laser is operated in the weak outcoupling regime producing a single spin component, classically quiet low divergence matter wave output beam with an inferred linewidth that can be a factor of 100 less than that achieved in the work of Hagley {\em et al.} \cite{comment1}. The ${\rm ^{87}Rb}$ Raman atom laser has a flux limit that we calculate to be a factor of 25 higher than that achievable with an RF outcoupled atom laser produced from the same source. As we will show, the output mode has a calculated divergence that can be a factor of at least 7 less in each transverse direction than any continuous (long pulse) atom laser demonstrated to date.  For our experiments, the limit for the beam brightness of a Raman outcoupled atom laser is more than three orders of magnitude greater than can be achieved with an RF output coupler. 

Our setup for producing BEC, the source for the atom laser, is unchanged from the apparatus  used in our previous studies of the RF continuous output-coupler \cite{n1}.  Briefly, we 
produce an ${\rm F=2, m_{F}=2}$  ${\rm ^{87}Rb}$ condensate, consisting of 
approximately $10^5$ atoms, via evaporation in a water-cooled QUIC 
magnetic trap \cite{esslinger} with a radial trapping frequency of
$\omega=2\pi \times260$\,Hz and an axial trapping frequency $\omega_z=2\pi \times20$\,Hz. 
We operate our trap at 
a bias field of ${\rm B_0=0.25}$\,G.  The trap runs at 12\,A (12\,V)  
generated from a low noise power supply.
The low power dissipation, and additional chilled water cooling, suppresses heating related drift of the magnetic trap bias 
field, allowing precise addressing of the condensate with resonant RF radiation or Raman beams (we measure a drift of significantly less than 0.7\,mG over 8 hours).  

The ${\rm F=2, m_{F}=2}$ atoms in the condensate are coupled to the ${\rm m_F}$= 0 state using two phase locked lasers. The coupling scheme is shown in figure 1a \cite{warning}.  Atoms coupled to the ${\rm m_F}$ = 0 state fall under gravity to produce the atom laser beam also shown in figure 1b.   The two phase locked optical beams that form the outcoupler are produced from a single 70\,mW diode laser that is split and sent through two separate phase locked AOMs, each in a double pass configuration. The frequency between the AOMs  corresponds to the  Zeeman plus kinetic energy difference between the initial and final states of the multi-photon Raman transition.   The beams are then coupled via a single mode, polarization maintaining optical fiber, directly to the BEC through a collimating lens and wave-plate, providing a maximum intensity of 250\,${\rm mW/cm^2}$ per beam.   The beams intersect at the condensate and are alligned in the plane of the weak axis and gravity and are separated by $\theta=45$ degrees in the vertical direction.  With the laser polarisations chosen appropriately the atoms acquire a momentum kick of ${\rm 4\hbar k \sin(\theta/2)}$ downwards.  The two coupling lasers, each with 60\,$\mu$W in roughly collimated 1\,mm beams (7.6\,${\rm mW/cm^2}$),  were applied to the condensate for 8.5\,ms.   The magnetic trap was then abruptly turned off and 2\,ms later the atoms were imaged to produce the data in  figure 1b.  
 \begin{figure}[t]
\centerline{\scalebox{.4}{\includegraphics{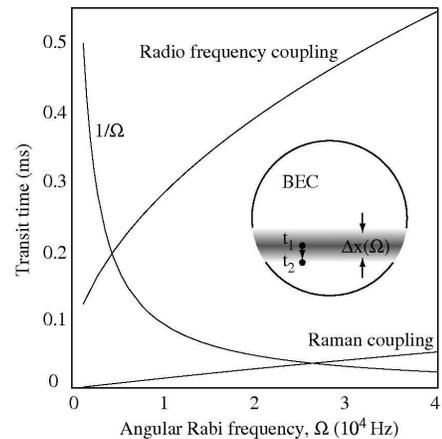}}}
\caption{A comparison of the transit time of atoms through the output coupling resonance $\Delta x$ for the RF and Raman output-couplers.  The inset shows a simple schematic of our model of continuous output coupling}
\end{figure}

In the following paragraphs, we show that the continuous Raman output coupler has the potential to surpass the output brightness \cite{definition} achievable in an RF atom laser by more than three orders of magnitude. The large momentum kick  imparted by the Raman lasers (up to ${\rm 4\hbar k}$ or a velocity of $\rm{\sim 2.35\,cm/s}$) boosts the output flux and decreases the transverse and longitudinal momentum spread of the atom laser.   We first consider the flux from a simple two state model of output-coupling (say $|F=1,m_F=-1\rangle$ to $|F=1,m_F=0\rangle$).  Applying a weak coupling between magnetically trapped and untrapped states leads to a localized output resonance within the BEC (figure 2 inset).  Gravity causes the condensate to sag (by $g/\omega^2$ where g is gravity and $\omega$ is the angular radial frequency of the magnetic trap) away from the magnetic field minimum, broadening the frequency resonance.

In the limit of continuous weak coupling in typical magnetic traps, the output region is composed of a roughly planar slice through the condensate  perpendicular to gravity \cite{bloch}.  The rate at which atoms are coupled between the fields is given by the angular Rabi frequency, $\Omega=g_F\mu_BB/(2 \hbar)$ for RF, where $g_F$ is the g-factor, $\mu_B$ is the Bohr magneton, B the magnetic field magnitude, and $\Omega=\Omega_1\Omega_2/(4\Delta)$ for Raman, where $\Omega_{1,2}$ are the single photon Rabi frequencies and $\Delta$ is the detuning from the appropriate resonance.  The characteristic frequency width of the coupling is the Rabi frequency.  For reasons of mathematical simplicity we arrange the center of the output-coupling resonance to coincide with the center of the condensate. We can then calculate the {\em characteristic spatial width} of the resonance, $\Delta x$, and hence a transit time for atoms to fall out of the resonance. The lower resonant half-width is found by considering the difference in resonant frequencies between the lower edge of the resonance at $x_l=g/\omega^2+\Delta x_l$ and the upper edge at $x_u=g/\omega^2$.   The magnetic field at any point in the vertical direction is given by $B=B_0+1/2B^{\prime\prime}x^2$ where $B^{\prime\prime}=\frac{m\omega^2}{g_F\mu_B}$ and m is the atomic mass, and the resonant frequency is $\omega_{rf}=g_F\mu_BB/\hbar$.  Thus the resonant width of the coupling is given by
\begin{equation}\label{reson}
\Delta\omega_{rf}=\frac{M\omega^2}{2\hbar}(x_l^2-x_u^2)=\frac{M\omega^2}{2\hbar}\left( \frac{2g\Delta x}{\omega^2} +{ \Delta x}^2\right).
\end{equation} 

This equation can be solved for the spatial width of the lower half of the output-coupling resonance as,
\begin{equation}
\Delta x_l=-\frac{g}{\omega^2}+\sqrt{\frac{g^2}{\omega^4}+\frac{\hbar\Omega}{M\omega^2}},
\end{equation}
where we have set $\Delta\omega_{rf}$ to the half-width of the frequency resonance, $\Omega/2$.  A similar expression can be derived for the upper half of the resonance to give a total spatial extent of $\Delta x= \Delta x_l+\Delta x_u$.
\begin{figure}[t]
\centerline{\scalebox{.4}{\includegraphics{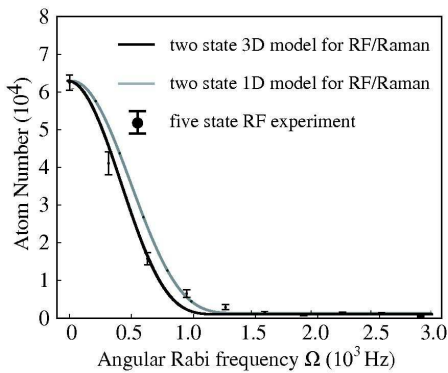}}}
\caption{A comparison of experiment and theory for the output flux of an RF atom laser.  Plotted is the number of atoms remaining trapped after 100\,ms of continuous output-coupling as a function of Rabi frequency. Our model for flux does differentiate between RF and Raman.}
\end{figure}

For an RF atom laser, an output-coupled atom is transferred from a trapped to an un-trapped state and is then accelerated out of the coupling region by gravity.   The output coupling will remain uncomplicated if the transit time of the atom through the coupling region, $\tau_{grav}=\sqrt{2\Delta x/g}$, is shorter than the Rabi flopping time \cite{n1}.  If the transit time exceeds the Rabi flopping time the atoms will be coupled back into the trapped Zeeman state (for this two state model) and the type of complex dynamics we have discussed previously will occur \cite{n1}.  Atoms output-coupled with a momentum kick, as in the Raman scheme, leave the interaction region more quickly, ($\tau_{kick} = \sqrt{(v_0/g)^2 + 2\Delta x/g} - v_0/g$ where $v_0$ is the velocity imparted to the atoms), and hence the homogeneous flux limit will be higher for this type of atom laser.  For comparison, the transit times of both an RF and a Raman output-coupler are plotted in figure 2 as a function of the coupling strength, which is parameterised by the Rabi frequency, $\Omega$.  The limit of weak out-coupling is given by the intersection of $1/\Omega$ with the transit time in figure 2. The two state RF atom laser satisfies the condition for weak out coupling and therefore classically quite operation for $\Omega < $5\,kHz. This result is consistent with the data we recorded in our investigation of flux and classical fluctuations in an RF atom laser \cite{comment2}. Based on this experimentally verified criterion, a Raman atom laser can be operated at a Rabi frequency of up to 25\,kHz, a factor of five higher than an RF atom laser.  Furthermore, in our Raman outcoupling experiment,  up to the maximum power available (250\,$mW/cm^2$) we find that the Raman atom laser {\em does not} exhibit a bound state \cite{jeffers}.  

We can extend this model to show that the Raman coupler can provide a significant increase in homogeneous output flux.     Integrating the condensate density  $|\Phi(x,y,z)|^2=1/U(\mu-V(x,y,z))$ (within the Thomas-Fermi approximation) over the output-coupling slice, $\Delta x$, we find the total number of atoms within the output coupling region, $N_{\Delta x}(N,\Omega)$.  Here $U=(4\pi\hbar^2 a)/m$ is the interaction strength and $a$ is the s-wave scattering length, $\mu=\frac{1}{2}(15a\hbar^2 m^{1/2}\omega_z \omega^2 N)^{2/5}$ is the chemical potential, with N the total number of atoms in the condensate and $V(x,y,z)=\frac{m}{2}(\omega^{2}(x^{2}+y^{2})+
\omega_{z}^{2}z^{2})$ is the trapping potential.  
The output flux of the atom laser can be approximated by multiplying the number of atoms in the coupling region by the Rabi frequency to give
\begin{equation}
 F=N_{\Delta x}(N,\Omega)\Omega.
 \end{equation}  
Figure 3 shows the numerical solution of the differential equation $d N/d t=-F$ for the number of atoms remaining in the condensate after 100\,ms of RF output-coupling and compare with our experiment.  We obtain excellent agreement with experiment {\em with no free parameters} in our model.  A one dimensional model of the same form, with the 1D condensate interaction term fixed by matching 1D and 3D chemical potentials is also in excellent agreement with the experiment.  A number of previous attempts to calculate the flux have found large discrepancies between 1D and 3D calculations and between theory and experiment \cite{aspect2}.  A previous experiment by Bloch et. al. on an RF atom laser \cite{bloch} measured an exponential decay of the condensate number with a rate  $1.2(2)\times10^{-5}\,s/\Omega^2$.  Our data gives a value of $3\times10^{-5}\,s/\Omega^2$.

Raman atom lasers can be expected to be significantly brighter than RF atom lasers because of their lower divergence  \cite{hagley}.  The repulsive mean-field experienced by the output atoms across the coupling resonance leads to a non-zero longitudinal and transverse velocity width of the atom laser beam. The transverse component of this velocity width leads to an undesirable divergence \cite{aspect} and mode quality \cite{kohl1,busch,kohl2,riou} of the RF atom laser.   Following \cite{busch}, we use a classical trajectories model to estimate the effect of the momentum kick of a Raman atom laser on the transverse and longitudinal velocity widths.   The results of this model are presented in figure 4 for the longitudinal and transverse velocities at the Thomas-Fermi edge of the condensate for the tightly confining axes of the magnetic trap.  The Raman atom laser gains a factor of 245 increase in brightness over the RF system due to a reduction of 7 in both transverse velocity widths and a factor of 5 in the longitudinal width.   
 
\begin{figure}[t]
\centerline{\scalebox{.35}{\includegraphics{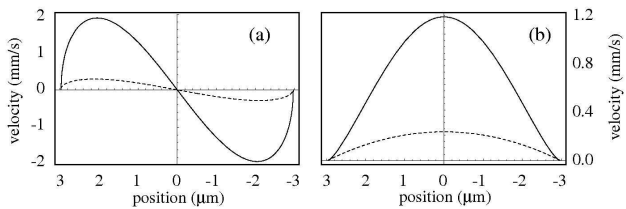}}}
\caption{Velocity distribution of the output coupled atoms after traversing the tightly confining longitudinal and transverse axes of the condensate, from an initial condition corresponding to a planar slice through the center of the condensate.  The velocity due to gravity and/or the kick has been subtracted to show only the velocity contribution from the condensate mean field.  (a) shows the transverse velocity distribution of the RF (solid) and Raman (dashed) atom laser, (b) the longitudinal distribution.  The Raman atom laser is operated with the maximum $4\hbar k$ kick in the direction of gravity.  Similar gains occur along the weakly confined transverse direction of the condensate.}
\end{figure}

From these considerations, we can estimate the {\em peak} flux of an atom laser and the maximum phase sensitivity in an interferometric measurement.  The maximum density obtainable in a $^{87}Rb$ condensate is clamped at approximately  $10^{14}$\,atoms/cm$^3$ by three body recombination.    Taking $\Omega$ to be the maximum allowable for the RF and Raman atom lasers (see figure 2) we find a peak flux of $1.4\times 10^8$ atoms/s and $4.2\times 10^{9}$ atoms/s respectively.  In making this estimate we have assumed realistically weak axial and radial trapping frequencies of 20 Hz.  The factor of five increase in Rabi frequency translates to a factor of 25 increase in peak flux for the Raman atom laser over the RF atom laser. In a shot noise limited measurement of phase the sensivity of an interferometric measurement of phase is $\pi/\sqrt{N}$ per square root of bandwidth where N is the total atom flux. Using the flux limit given above for Raman outcoupling, the maximum sensitivity of a Raman atom laser based intreferometric phase measurement is approximately $10^{-5}$ rad/$\sqrt{Hz}$, a factor of five times more sensitive than that achievable with an RF atom laser. This will be the case whether the system is unpumped or perhaps pumped in future experiments. The only way to exceed this sensitivity is to increase the flux by using non-state changing out coupling, to increase the density past the current limit set by three body recombination, or to decrease the shot noise through squeezing. The first option appears to be the most likely path in the immediate future. It is important to note however, that the Heisenberg limit sets a sensitivity limit of $\pi$/N rad/Hz or $2\times 10^{-10}$\,rad/Hz if we use the peak flux for a $^{87}$Rb Raman laser. It is interesting to consider the possibility that very strong squeezing is possible due to high non linearities in an atomic system and this may be very important in increasing the sensitivity of atom laser based precision measurment in future experiments. Finally we note that a condensate source of lighter atoms will produce vastly higher homogeneous output flux from a Raman output coupler.  For example, the recoil velocity of hydrogen, from a ${\rm 2\hbar k}$ momentum kick, is 330\,cm/s at an excitation wavelength of 243\,nm.


\begin{thebibliography}{0}
\bibitem{comment1} The energy linewidth of a continuous atom laser in the weak coupling limit is fourier limited by the output coupling duration.
\bibitem{warning} It is important to note that the usual theoretical lambda Raman scheme of $\Delta m=2$ transitions driven by $\sigma_+$, $\sigma_-$ polarisations cannot be implemented experimentally if the one photon detuning is large enough to suppress spontaneous emission.
\bibitem{definition}Brightness is defined here as the integrated flux of atoms per source size divided
by the velocity spread in each dimension.
\bibitem{comment2}For a five state atom laser the flux limit will be lower than the ideal two state system presented here because interactions with the anti-trapped Zeeman states add extra classical fluctuations.
\end{thebibliography}
\end{document}